\documentclass[pra, twocolumn, superscriptaddress, preprintnumbers, floatfix]{revtex4-1}
\usepackage{float}
\usepackage{amsmath}
\usepackage{amssymb}
\usepackage{graphicx}
\usepackage{bm}
\usepackage{amsfonts}
\usepackage{mathtools}
\usepackage{color}
\usepackage{wasysym}
\usepackage{latexsym}
\def\ulamek#1#2{\mbox{\normalfont$\frac{#1}{#2}$}}

\begin{document}
%\makeatletter

\title{Relativistic Wave Equations: An Operational Approach}

\author{G. Dattoli}
\email{dattoli@frascati.enea.it}

\author{E. Sabia}
\email{sabia@frascati.enea.it}
\affiliation{ENEA - Centro Ricerche Frascati, via E. Fermi, 45, IT 00044 Frascati
(Roma), Italy}

\author{K. G\'{o}rska}
\email{katarzyna.gorska@ifj.edu.pl}

\author{A. Horzela}
\email{andrzej.horzela@ifj.edu.pl}
\affiliation{H Niewodnicza\'{n}ski Institute of Nuclear Physics, Polish Academy of
Sciences, ul. Eljasza-Radzikowskiego 152, PL-31342 Krak\'{o}w, Poland}

\author{K. A. Penson}
\email{penson@lptl.jussieu.fr}
\affiliation{{Sorbonne Universit\'{e}s}, Universit\'e Pierre et Marie Curie (Paris 06), CNRS UMR 7600\\
Laboratoire de Physique Th\'eorique de la Mati\`{e}re Condens\'{e}e (LPTMC),\\
Tour 13 - 5i\`{e}me \'et., B.C. 121, 4 pl. Jussieu, F 75252 Paris Cedex 05, France\vspace{2mm}}

\begin{abstract}
The use of operator methods of algebraic nature is shown to be a very powerful tool to deal with different forms of relativistic wave equations. The methods provide either exact or approximate solutions for various forms of differential equations, such as relativistic Schr\"{o}dinger, Klein-Gordon and Dirac. We discuss the free particle hypotheses and those relevant to particles subject to non-trivial potentials. In the latter case we will show how the proposed method leads to easily implementable numerical algorithms. 
\end{abstract}
\maketitle

\section{Introduction}

Operation techniques embedded with the formalism of the evolution operator  have played a central role in the study of time dependent problems in quantum mechanics. The associated technicalities have attracted the attention of physicists   \cite{Baym} and mathematicians \cite{Wilcox}, who provided tools of crucial importance to deal with the study of the time dependent Schr\"{o}dinger equation. Expansions of the evolution operator like those developed by Stueckelberg,  Feynman and Dyson  on the physicists' side and by Magnus  and Fer on the other have paved the way to the formalism underlying the modern diagrammatic procedures \cite{Dattoli88, Dattoli97, Blanes}.
Evolution problems in relativistic quantum mechanics have raised new questions mainly associated with the pseudo-differential nature of the involved operators.  These methods, neither popular nor well appreciated by physicists in the past, are now getting increasing attention  for a more sound formulation of the underlying mathematical foundations and for the wealth of phenomenology they can describe \cite{Glaeske}. In particular, the operational approach has been shown to be effective in solving pseudo-differential equations, as well as those containing fractional derivatives and related to anomalous transport phenomena. The latter include not only the  so-called fractional dynamics adopted to  investigate anomalous diffusion and similar effects in complex physical systems \cite{Eliazar} but appear useful to model problems in life sciences, engineering, economy, and even studies of human mobility \cite{Mark, Klaffter, Barndorff, Stollenwerk, Gonzalez, Ramos, Sims}.

In this paper we will focus our attention on the equations of the relativistic quantum mechanics. Our aim is to push forward previous investigations of solutions to the relativistic evolution equation involving fractional derivatives, namely the relativistic Schr\"{o}dinger (called also Salpeter's) equation \cite{Babusci11, Babusci13,KowalskiRembielinski}, and to compare them with operationally obtained solutions to the Dirac-type and Klein-Gordon equations. The main motivation of this research is that fundamental relativistic wave equations give results whose interpretation based on Hamiltonian mechanics concepts, as well as explanations of arising paradoxes (like Zitterbewegung or the Klein paradox), still remains controversial \cite{GreinerBjoerken} and need clarification. In our opinion the use of fractional derivatives or of pseudo-differential operators, to treat relativistic quantum problems, allows a more complete understanding of longstanding problems associated with  nonlocal evolution problems. These problems, once considered academic, have in recent years acquired a more concrete flavour, mainly in the context of the so-called quantum simulation, namely the possibility of "simulating" quantum nonrelativistic effects in condensed matter and atomic physics \cite{Strange}.

We begin with recalling results coming from the study of the relativistic heat equation (as proposed in \cite{Babusci11}) whose  solutions provide us an example of the method. The Section III is devoted to the analysis of the free relativistic Schr\"{o}dinger equation and unexpected behaviour of its solutions. In the Section IV we consider the Dirac factorization of the relativistic Schr\"{o}dinger equation and compare so obtained results with those of Section III. The Klein-Gordon equation, together with the relation of its solutions to the Dirac case, is studied in the Section V, while in the Section VI we present the analysis of the simplest ``interacting model'', namely the relativistic particle (Schr\"{o}dinger's and Dirac's) driven by a linear potential. In what follows we restrict ourselves to 1+1 spacetime models. It simplifies the presentation and preserves readability of the results but there is no reason to assume that general properties of solutions will not be shared with the realistic 1+3 situation. 

\ \\ \ \\

\section{The relativistic heat equation}

In Ref. \cite{Babusci11} the following equation has been defined as the relativistic heat equation 
\begin{align}\label{EQ1}
\partial_{t}F(x,t) &= -\sqrt{d^{2}-k^{2}\partial_{x}^{2}}F(x,t),\\
F(x,0) & = f(x), \label{EQ2}
\end{align}
where  $t$ has the dimension of time, $d$ and $k$ have the dimensions of the inverse time and of a length divided by time, respectively. Introducing dimensionless variables $\tau = d t$ and $\xi = x d/k$ we can cast Eq. \eqref{EQ1} in the form
\begin{align}\label{EQ3}
\partial_{\tau}F(\xi,\tau) & = -\sqrt{1-\partial_{\xi}^{2}} F(\xi,\tau).
\end{align}
From the mathematical point of view Eqs. \eqref{EQ1} and \eqref{EQ3} are evolution equations with the distinctive feature that they contain the square root of a differential operator. The formal solution of Eq. \eqref{EQ3} can be written as
\begin{equation}\label{EQ4}
F(\xi,\tau) = e^{-\tau\sqrt{1-\partial_{\xi}^{2}}}f(\xi).
\end{equation}

Albeit the use of the exponential operators is quite common, the handling of exponentials containing the square root of a differential operator (as well as higher order roots) have been rarely treated. Nevertheless, some analytical tools have been developed for this purpose in the past. In the specific case of Eq. \eqref{EQ4} a suitable representation, which allows evaluation of the action of "square rooted" operator on the function $f(\xi)$, can be realized through the Doetsch formula \cite{Doetsch}. The latter allows explicit evaluation of the r.h.s. of Eq.\eqref{EQ4} according to the following integral transform
\begin{align} \label{EQ5}
F(\xi,\tau) &=\left[ \int_{0}^{\infty} g_{1/2}(\eta)e^{-\eta\tau^{2}(1-\partial_{\xi}^{2})}d\eta\right] f(\xi), \\
& = \int_{0}^{\infty} g_{1/2}(\eta)e^{-\eta\tau^{2}} D(\xi, \eta\tau^{2}) d\eta. \label{EQ6}
\end{align}
where $D(\xi, \eta\tau^{2})= e^{\eta\tau^{2}\partial_{\xi}^{2}}\, f(\xi)$. Note two peculiar features of the Eq. \eqref{EQ6}:
\begin{itemize}
\item{
Solution of the relativistic heat equation Eq. \eqref{EQ1} depends, via an appropriate transform,
on the solution of the ``classical'' diffusion problem as $D(\xi, \tau)$ is the solution of the ordinary heat equation
\begin{align}\label{EQ7}
\partial_{\tau} D(\xi, \tau) &= \partial_{\xi}^{2} D(\xi, \tau), \\
D(\xi, 0) &= f(\xi), \label{EQ8} 
\end{align}
usually expressible in terms of the Gauss-Weierstrass transform
\begin{equation}\label{EQ9}
D(\xi, \tau) = e^{\tau \partial^{2}_{\xi}} f(\xi) = \frac{1}{2\sqrt{\pi\tau}}\intop_{-\infty}^{\infty}e^{-\frac{\left(\xi-\sigma\right)^{2}}{4\tau}} f(\sigma)d\sigma.
\end{equation}
or written, using the Fourier transform, as
\begin{equation}\label{EQ10}
D(\xi, \tau) = \frac{1}{\sqrt{2\pi}} \int_{-\infty}^{\infty} e^{i q \xi} e^{\tau q^{2}} \tilde{f}(q) dq,
\end{equation}
with $\tilde{f}(q) = (2\pi)^{-1/2} \int_{-\infty}^{\infty} e^{-iq  \xi} f(\xi) d\xi$.
}
\item{
The integral kernel 
\begin{equation}\label{EQ11}
g_{1/2}(\eta) = \frac{1}{2\sqrt{\pi}\eta^{3/2}} e^{-1/(4\eta)} 
\end{equation}
is a generic example of the so-called L\'{e}vy stable distributions or $\alpha$-stable distributions. The latter denoted as $g_{\alpha}(\eta)$ are for $\eta\ge  0$, and for $0 < \alpha < 1$, given \textit{uniquely} by the inverse Laplace transform of $e^{-p^{\alpha}}$, i.e. $e^{-p^{\alpha}} = \int_{0}^{\infty} e^{-p\eta} g_{\alpha}(\eta) d\eta$ see \cite{Montroll, Penson, Gorska}}.
\end{itemize}

Let us now consider a solution of the relativistic heat equation for two specific initial functions. First, we assume that the initial condition of Eq. \eqref{EQ8} is a Gaussian: 
\begin{equation}\label{EQ12}
f_{1}(\beta; \xi) = \frac{1}{2\sqrt{\pi \beta}} \exp\left(\!-\frac{\xi^{2}}{4\beta}\!\right)=\frac{1}{2\pi}\int_{-\infty}^{\infty}e^{-\beta k^{2} + i k \xi} dk, 
\end{equation}
where $\beta>0$. We can therefore write the relevant solution as
\begin{align}\nonumber
D_{1}(\beta; \xi, \tau) &= \frac{1}{2\sqrt{\pi (\beta + \tau)}} \exp\left[-\frac{\xi^{2}}{4(\beta + \tau)}\right] \\
&= \frac{1}{2\pi} \int_{-\infty}^{\infty} e^{-(\beta +\tau) k^{2}} e^{i k \xi} dk,\label{EQ13}
\end{align}
recognized as the so-called Glaisher identity \cite{Dattoli}. Eq.~\eqref{EQ13}, if substituted into Eq. \eqref{EQ6}, leads to
\begin{equation}\label{EQ14}
F_{1}(\beta; \xi, \tau) = \frac{1}{2\pi} \int_{-\infty}^{\infty} e^{-\beta k^{2}} e^{-\tau \sqrt{1+k^{2}}} e^{i k \xi} dk,
\end{equation}
the same as the result of formal manipulation on Eqs.\eqref{EQ4} and \eqref{EQ12}.

The second example of the initial condition under consideration  is
\begin{equation}\label{EQ15}
f_{2}(\beta; \xi) = \frac{1}{2K_{1}(\beta)} \int_{-\infty}^{\infty} e^{-\beta \sqrt{1+k^{2}} + ik\xi} dk,
\end{equation}
where $\beta > 0$, $K_{1}(\beta)$ is the modified Bessel function of the second kind, i.e. the Macdonald function \cite{Andrews}. Eq. \eqref{EQ15} is essentially the Fourier transform of
\begin{equation}\label{EQ16}
\tilde{f}_{2}(\beta; k) = \frac{1}{2K_{1}(\beta)} e^{- \beta \sqrt{1+k^{2}}}
\end{equation}
normalized in the ${\cal L}^{1}$-norm: $\int_{-\infty}^{\infty} \tilde{f}_{2}(\beta; k) dk~=~1$. The solution of the relativistic heat equation with Eq.  \eqref{EQ15} as initial function, calculated as shown before, reads
\begin{align}\label{EQ17}
F_{2}(\beta; \xi, \tau) &= \frac{e^{-\tau\sqrt{1-\partial_{\xi}^{2}}}}{2K_{1}(\beta)} \int_{-\infty}^{\infty} e^{-\beta\sqrt{1+k^{2}} + ik\xi} dk \nonumber \\
& = \frac{1}{2K_{1}(\beta)} \int_{-\infty}^{\infty} e^{- (\beta+\tau) \sqrt{1+k^{2}} + ik\xi} dk.
\end{align}
The use of the identity (see Refs. \cite{Andrews, Prudnikov})
\begin{equation}\label{EQ18}
\int_{-\infty}^{\infty}\!\! e^{-a\sqrt{x^{2}+q^{2}}} \cos(b x)\, dx = \frac{2 a q}{\sqrt{a^{2}+b^{2}}} K_{1}(q \sqrt{a^{2}+b^{2}})
\end{equation}
allows to cast Eq. \eqref{EQ17} in the form
\begin{equation}\label{EQ19}
F_{2}(\beta; \xi, \tau) = \frac{\beta+\tau}{2K_{1}(\beta)}\, \frac{K_{1}(\sqrt{(\beta+\tau)^{2} + \xi^{2}})}{\sqrt{(\beta+\tau)^{2}+\xi^{2}}},
\end{equation}
which may be viewed as the identity
\begin{align}\label{EQ20}
&e^{-\tau\sqrt{1-\partial_{\xi}^{2}}}\, \frac{\beta}{2K_{1}(\beta)}\, \frac{K_{1}(\sqrt{\beta^{2}+\xi^{2}})}{\sqrt{\beta^{2}+\xi^{2}}} \nonumber \\
&\qquad\qquad\quad = \frac{\beta+\tau}{2K_{1}(\beta)}\, \frac{K_{1}(\sqrt{(\beta+\tau)^{2}+\xi^{2}})}{\sqrt{(\beta+\tau)^{2}+\xi^{2}}},
\end{align}
and interpreted as a generalization of the Glaisher formula for the Macdonald function.

Concluding this section we would like to note that naming Eq.\eqref{EQ1} as relativistic heat equation may lead to some misconception and should be used carefully, or even in quotes. The concept of relativistic diffusion needs clarification at least from the mathematical point of view. Namely Eq.\eqref{EQ1} does not represent a diffusion process in the strict sense because the norm of the relevant solution is not preserved during the evolution and decays with $e^{-\tau}$. Eqs. \eqref{EQ14} and \eqref{EQ19} are properly normalized at any time if multiplied by $e^{\tau}$ and $K_{1}(\beta)e^{\beta+\tau}/\pi$, respectively. In such a case they may be considered as the evolution of a distribution which satisfies the equation $\partial_{\tau}F_{i}(\xi, \tau) = -[(1 - \partial_{\xi}^{2})^{\frac{1}{2}} - 1] F_{i}(\xi, \tau)$, $i=1, 2$ and exhibits the correct nonrelativistic limit in the sense that $-[(1-\partial_{\xi}^{2})^{\frac{1}{2}} - 1] \simeq \ulamek{1}{2}\partial_{\xi}^{2} - \ulamek{1}{8}\partial_{\xi}^{4}+\ldots$. \

\section{The Relativistic Schr\"{o}dinger equation (Salpeter equation)}

The example of relativistic evolution, which will be considered here, is the free relativistic Schr\"{o}dinger equation, called also the Salpeter equation, which has the form
\begin{equation}\label{EQ21}
i \hbar\, \partial_{t} \Psi = c \sqrt{{\hat p}^{2} + m_{0}^{2} c^{2}} \Psi.
\end{equation}
Written in dimensionless variables it reads 
\begin{equation}\label{EQ22}
i \partial_{\tau} \tilde{\Psi}(q, \tau) = \sqrt{1 + {\hat q}^{2}}\, \tilde{\Psi}(q, \tau),
\end{equation}
where $\tau = ct/\lambda_{c}$, ${\hat q} = \lambda_{c} {\hat p}/\hbar$, and $\lambda_{c} = \hbar/(mc)$, and then 
\begin{align}\label{EQ23}
i \partial_{\tau} \Psi(\xi, \tau) &= \sqrt{1-\partial_{\xi}^{2}}\, \Psi(\xi, \tau), \nonumber\\
\left.\Psi(\xi, \tau)\right|_{\tau=0} &= \Psi_{0}(\xi).
\end{align}
with ${\hat q}=i\partial_{\xi}$,  $\xi = x/\lambda_{c}$, originating from the standard quantization rule ${\hat p}=i\hbar\partial_{x}$.

In analogy to the Eq. \eqref{EQ6} we can express the formal solution of Eq. \eqref{EQ23} in terms of a Gaussian wave function $D(\xi, \tau)$ and the L\'{e}vy distribution $g_{1/2}(\eta)$
\begin{align}\label{EQ24}
\Psi(\xi, \tau) &= e^{-i\tau\sqrt{1-\partial^{2}_{\xi}}}\, \Psi_{0}(\xi) \nonumber \\
&= \int_{0}^{\infty}\!\! g_{1/2}(\eta) e^{\eta \tau^{2}} D(\xi, - \eta \tau^{2}) d\eta.
\end{align}
Taking $D(\xi, -\eta\tau^{2})$ in the Fourier space (see Eq. \eqref{EQ10} with appropriately changed variables) we can show that Eq. \eqref{EQ24} is equivalent to the solution of Eq. \eqref{EQ22}. It is seen from
\begin{align}\label{EQ25}
\Psi(\xi, \tau) &= \int_{0}^{\infty}\!\! g_{1/2}(\eta) e^{\eta\tau^{2}} \left[\int_{-\infty}^{\infty}\!\! e^{i q \xi} e^{\eta\tau^{2} q^{2}} \tilde{\Psi}_{0}(q) \frac{dq}{\sqrt{2\pi}}\right] d\eta \nonumber \\
& = \int_{-\infty}^{\infty}\!\! e^{i q \xi} \left[\int_{0}^{\infty}\!\! g_{1/2}(\eta) e^{\eta\tau^{2}(1+q^{2})} d\eta\right] \tilde{\Psi}_{0}(q) \frac{dq}{\sqrt{2\pi}} \nonumber \\
& = \frac{1}{\sqrt{2\pi}} \int_{-\infty}^{\infty}\!\! e^{i q \xi} e^{-i \tau \sqrt{1+q^{2}}}\, \tilde{\Psi}_{0}(q) dq.
\end{align}
The inverse Fourier transform of Eq. \eqref{EQ25} defines $\tilde{\Psi}(q, \tau)$ which is a formal solution of Eq. \eqref{EQ22}.

Let us now solve the relativistic Schr\"{o}dinger equation for an initial Gaussian packet given in Eq. \eqref{EQ12}, i.e. for the initial condition $e^{-\beta q^{2}}$ taken in Eq. \eqref{EQ25}. Thus we have
\begin{equation}\label{EQ26}
\Psi_{1}(\beta; \xi, \tau) = \frac{N}{\sqrt{2\pi}} \int_{-\infty}^{\infty} e^{-\beta q^{2} - i \tau \sqrt{1+ q^{2}} + i q \xi} dq
\end{equation}
 with the normalization constant $N = [2\beta/(\pi\lambda^{2}_{c})]^{1/4}$ calculated in $\mathcal{L}^{2}[\mathbb{C}, dx]$ space.
 
\begin{figure}[!h]
\includegraphics[scale=0.4]{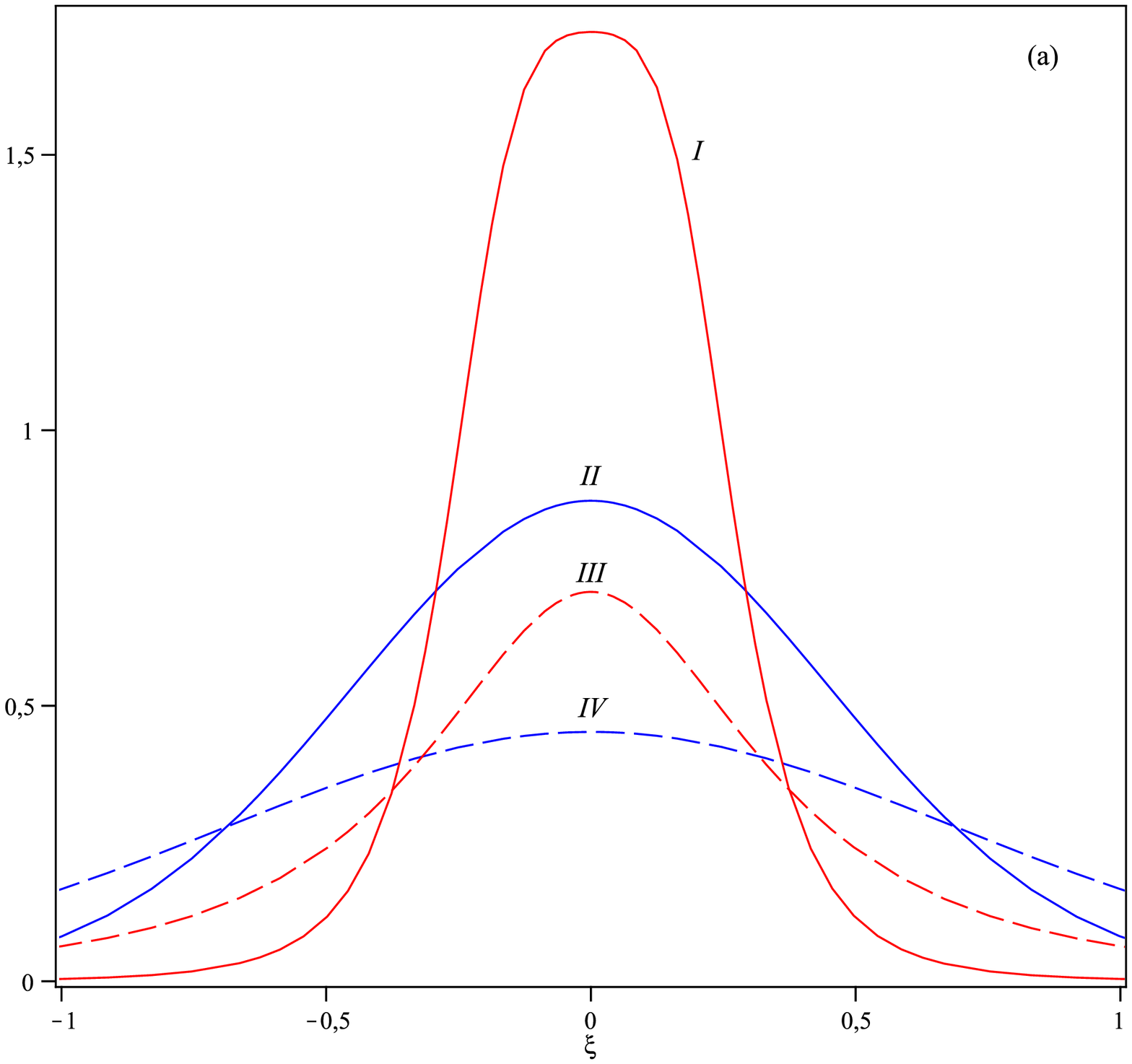}
\includegraphics[scale=0.4]{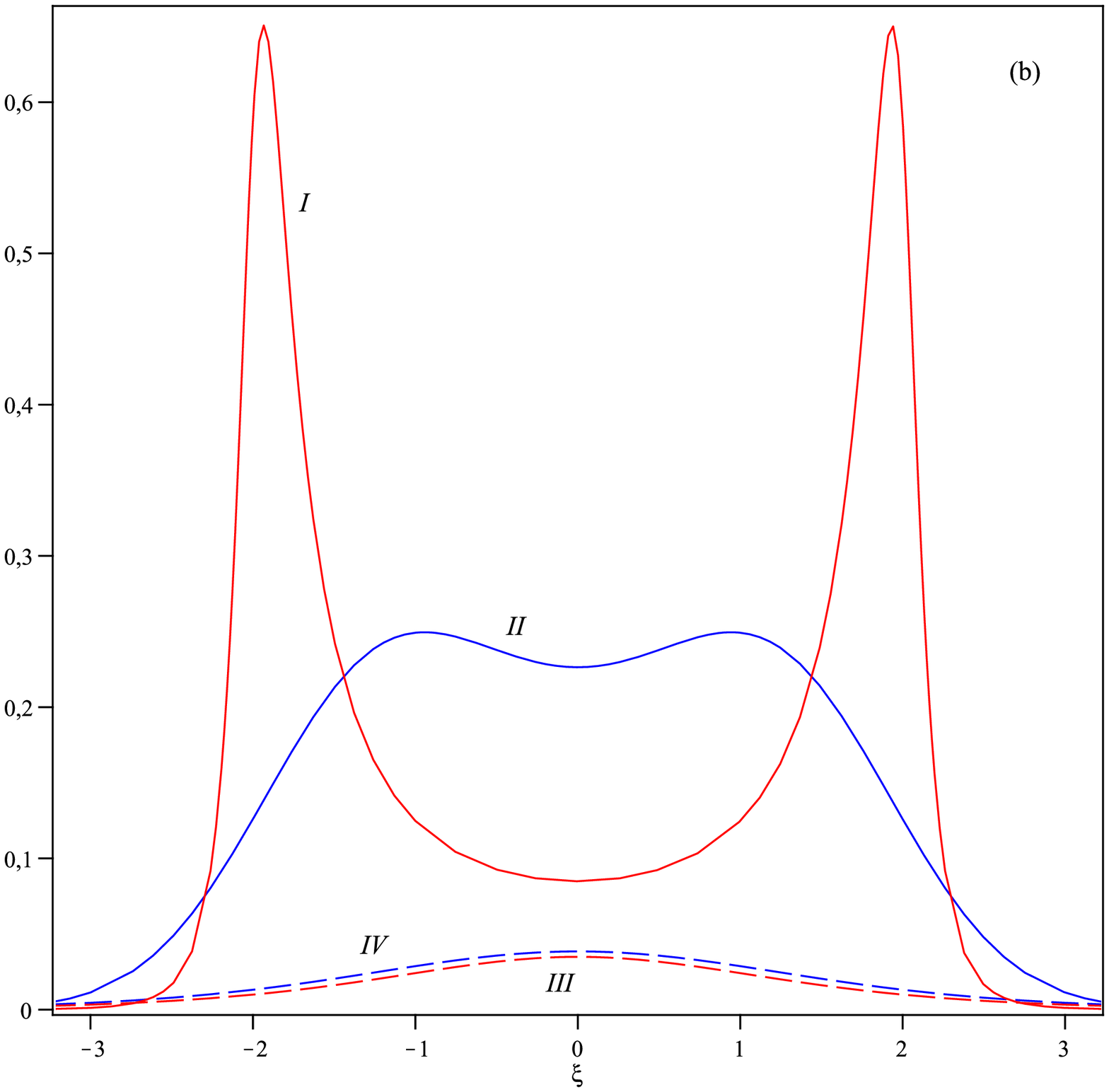}
\caption{\label{fig3}   (a): Comparison, for $\beta = 0.2$ and $\tau = 0.2$, of the relativistic solutions $|\Psi_{i}(\beta; \xi, \tau)|$, $i = 1, 2$ (Eqs. (27) and (26) for lines I and II, respectively) and nonrelativistic solutions $F_{i}(\beta; \xi, \tau)$, $i = 1, 2$ (Eq. (19) and (14) for lines III and IV, respectively).  (b): the same as in Fig. 1(a) but for $\beta = 0.2$ and $\tau = 2$.}
\end{figure}

A time-dependent solution of Eq. \eqref{EQ23} obtained along the same lines as before for $\tilde{\Psi}_{0}(q) =  \frac{1}{2K_{1}(2\beta)}e^{-\beta\sqrt{1+q^{2}}}$ reconstructs the result presented in \cite{KowalskiRembielinski}, see Eq. (4.18) there. Namely, we get
\begin{align}\label{EQ27}
\Psi_{2}(\beta; \xi,\tau) &= \frac{\beta e^{- i \tau \sqrt{1-\partial_{\xi}^{2}}}}{\sqrt{ \pi K_{1}(2 \beta)}}\, \frac{K_{1}(\sqrt{\beta^{2}+\xi^{2}})}{\sqrt{\beta^{2}+\xi^{2}}} \nonumber\\
&= \frac{\beta + i\tau}{\sqrt{ \pi K_{1}(2 \beta)}}\, \frac{K_{1}(\sqrt{(\beta+i\tau)^{2}+\xi^{2}})}{\sqrt{(\beta+i\tau)^{2}+\xi^{2}}}.
\end{align}
A comparison between nonrelativistic and relativistic solutions reveals interesting behaviour illustrated in Fig. \ref{fig3}a) and \ref{fig3}b). The snapshots, taken at different times, show that the wave packet spreading is ``slower'' for the relativistic case than for its nonrelativistic counterpart. Furthermore, the relativistic solutions reveal, at longer times, the wave function deformation determining their departure from the Gaussian-like initial form. 
Numerical calculations of $\Psi_{i}(\beta; \xi, \tau)$  done for various values of $\beta$ and $\tau$ show that there exists a non-trivial relation between $\beta$ and $\tau$ which implies an appearance, or non-appearence, of two peaks. Two peaks evolution pattern appears for reasonable values of $\tau$ if $\beta$ is small enough; if not, the evolution follows the standard spreading. As a quantity indicating what is going on for a given $\tau$ the value of the second derivative $D^{(2)}_{i}(\beta, \tau) =\left[\partial^{2}_{\xi}|\Psi_{i}(\beta; \xi, \tau)|^{2}\right]_{\xi=0}$, $i=1, 2$, may be used. For $\tau=0$ it is evident that    $D^{(2)}_{i}(\beta, 0)<0$ for both Eqs. \eqref{EQ26} and \eqref{EQ27}. But if $\tau$ grows it may happen that  $D^{(2)}_{i}(\beta, \tau)$ changes its sign, which means that the point $\xi=0$ becomes local minimum. This depends on the parameter $\beta$, measuring the  width of the initial wave packet, i.e. the initial localization, as well. From numerical calculation of $D^{(2)}_{i}(\beta, \tau) = \left[\partial^{2}_{\xi}|\Psi_{i}(\beta; \xi, \tau)|^{2}\right]_{\xi=0}\!\!\simeq 0$, $i=1, 2$, for fixed $\tau$ we can find that there exist $\beta_{c}$ such that for $\beta > \beta_{c}$ we can observe two peaks. For instance, $D^{(2)}_{1}(\beta, 7) \simeq 0$ for $\beta_{c} = 0.689$ and $D^{(2)}_{2}(\beta, 7) \simeq 0$ for $\beta_{c} = 1.411$. All this allows us to make a conjecture: two-peaks evolution pattern is possible only if the initial wave packet is localized sufficiently strongly, $\beta<\beta_{c}$ and the ``critical'' localization $\beta_{c}$ is related to the Compton wavelenght of the particle.

\begin{figure}[!h]
\includegraphics[scale=0.44]{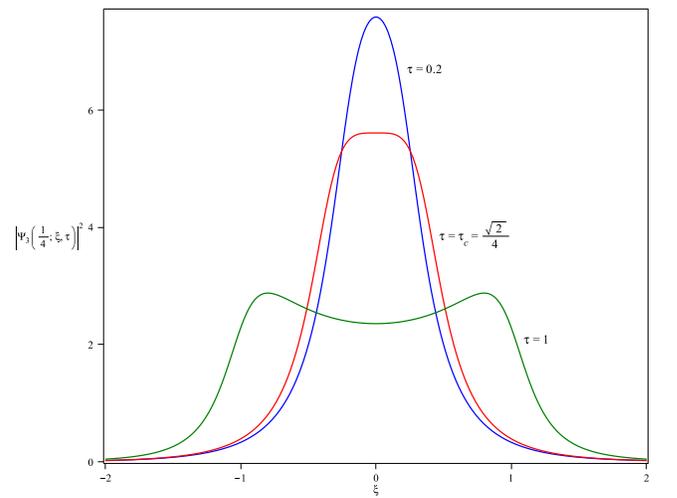}
\caption{\label{rys2} The three various behaviours of $|\Psi_{3}(\beta; \xi, \tau)|^{2}$ for $\beta = 1/4$ and $\tau = 0.2$ (blue line), $\tau = \tau_{c} = \sqrt{2}/4$ (red line), and $\tau = 1$ (green line).}
\end{figure}

To justify the previous assumption we will consider another solution of Eq. \eqref{EQ23}, which is obtained for the initial condition defined in the Fourier space as follows
\begin{equation}\label{EQ28}
\tilde{\Psi}^{(3)}_{0}(\beta; q) = \frac{N |q|}{(\sqrt{1+q^{2}} - 1)^{1/2}} \frac{e^{-\beta\sqrt{1+q^{2}}}}{\sqrt{1+q^{2}}}. 
\end{equation}
After a simple calculation, applying Eq. (2.5.39.6) on p. 456 of \cite{Prudnikov}, Eq. \eqref{EQ25}, with the initial condition Eq. \eqref{EQ28}, reads
\begin{align}\label{EQ29}
\Psi_{3}(\beta; \xi, \tau) = N\, (\sqrt{a^{2} + \xi^{2}} + a)^{1/2}\,  \frac{e^{-\sqrt{a^{2} + \xi^{2}}}}{\sqrt{a^{2} + \xi^{2}}},
\end{align}
where $a = \beta + i\tau$. The normalization constant $N$ calculated in $\mathcal{L}^{2}$ is given as $N = \sqrt{\pi/\lambda_{c}}[K_{0}(2\beta) + \ulamek{\pi}{2} - \pi\beta K_{0}(2\beta) \mathbf{L}_{-1}(2\beta) - \pi\beta K_{1}(2\beta) \mathbf{L}_{0}(2\beta)]^{-1/2}$, where $\mathbf{L}_{\nu}(u)$ is the modified Struve function. Now we can find the exact relation between $\beta$ and $\tau$ for which two peaks appear. Namely, 
\begin{align}\label{EQ30}
D^{(2)}_{3}(\beta, \tau) &= \frac{N}{2\sqrt{\pi}} \frac{e^{-2\beta}}{(\beta^{2} + \tau^{2})^{5/2}} \nonumber \\
&\times [\tau^{2}(3-4\beta) - \beta^{2}(3+4\beta)],
\end{align}
from which it is immediately seen that
\begin{itemize}
\item{ $D^{(2)}_{3}(\beta, 0)< 0$ for $\tau=0$,}
\item{ if $\beta\ge 3/4$ then $D^{(2)}_{3}(\beta, \tau)< 0$ for all $\tau$,}
\item{ if $\beta < 3/4$ then $D^{(2)}_{3}(\beta, \tau)$ becomes positive for  $\tau > \tau_{c} = (\beta^{2}(3+4\beta))/(3-4\beta)$,} 
\end{itemize} 
The behaviour of $|\Psi_{3}(\beta; \xi, \tau)|^{2}$ for $\beta = 1/4$, for which $\tau_{c} = \sqrt{2}/4$, are illustrated in Fig. \ref{rys2} in which the case $\tau < \tau_{c}$ is marked by blue line, $\tau = \tau_{c}$ is denoted by red line, and $\tau > \tau_{c}$ is green line.

Let us now come back to the consideration in dimensional coordinates $x$ and $t$. According to the choice of the normalization of variables the all initial wave function is localized within the Compton wavelength of the particle itself. For example, Eq. \eqref{EQ27} expressed in dimensional form reads
\begin{equation}\label{EQ31}
\Psi_{2}(x, t) = \frac{\beta\lambda_{c} + i c t}{\sqrt{\lambda_{c}\pi K_{1}(2\beta)}} \frac{K_{1}(\ulamek{1}{\lambda_{c}}\sqrt{(\beta\lambda_{c} + i c t)^{2} + x^{2}})}{\sqrt{(\beta\lambda_{c} + i c t)^{2} + x^{2}}}.
\end{equation}
The role of the parameter $\beta$ appearing in Eq. \eqref{EQ31} is not trivial.  It is an arbitrary constant associated with the intrinsic nonlocality of the problem under study and it defines the scale of delocalization of the particle within its Compton wavelength.

The wave packets whose evolution has been reported in Figs. \ref{fig3} are initially localized within a Compton wave-length. In the case of electronic neutrino assuming a mass of $2\, eV$ we are talking about an uncertainty in position of the order of few microns. The case of massless particles could be obtained from Eq. \eqref{EQ31} by going to the limit $\lambda_{c}\to\infty$. However keeping such a limit without any further caution does not provide physical results. Eq. \eqref{EQ31} yields indeed a completely delocalized function unless we assume that in such a limit $\beta\lambda_{c}$ remains a constant. Since $\beta$ is an arbitrary scale factor, we can define $b = \beta\lambda_{c}$. By such a redefinition the dependence on the Compton wave-length is a physical dependence and no more a scale factor. We can keep the limit safely and using the well-known asymptotic property $K_{1}(z)\vert_{z\to\infty}\simeq 1/z$, we obtain the same result as quoted in Ref. \cite{Kowalski}:
\begin{equation}\label{EQ32}
\Psi_{2}(x, t) = \sqrt{\frac{2b}{\pi}} \frac{b + i c t}{(b + i c t)^{2} + x^{2}}. 
\end{equation}
It is now worth to recast Eq. \eqref{EQ32} in the form 
\begin{equation}\label{EQ33}
\Psi(x, t) = \sqrt{\frac{b}{2\pi}}\left[\frac{1}{b + i(x + c t)} + \frac{1}{b - i(x - c t)}\right]
\end{equation}
and note that it can be viewed as the solution of the equation \cite{Kowalski}
\begin{align}\label{EQ34}
i\partial_{t}\Psi &= c\sqrt{-\partial_{x}^{2}}\, \Psi,  \\
\Psi(x, 0) &= \sqrt{\frac{b}{2\pi}} \left(\frac{1}{b + ix} + \frac{1}{b - ix}\right), \label{EQ35}
\end{align}
which is  the massless limit of the relativistic Schr\"{o}dinger equation. Eq. \eqref{EQ34} does not depend explicitly on the Planck constant; such a dependence is however implicitly contained in the arbitrary constant $b$. We find also that the r.m.s. value of the particle position is indeed provided by $\sqrt{\langle x(t)^{2}\rangle} = \sqrt{b^{2} + (c t)^{2}}$ which is consistent with the constraint $\sqrt{\langle p(t)^{2}\rangle} > \hbar/(2\sqrt{b^{2} + (c t)^{2}})$ (for additional comments see Ref. \cite{Babusci11}). The emergence of two peaks, when the particle is localized according to the previous discussion, is a distinctive feature of the nonlocal nature inherent to the  relativistic Schr\"{o}dinger equation. This aspect of the problem may acquire a deeper physical consistence when studied in the presence of a potential, as will be discussed in a forthcoming paper.

\section{The Dirac Factorization and the one dimensional Quantum Relativistic
Equation}

This section is devoted to the Dirac factorization of the one dimensional relativistic Schr\"{o}dinger equation. This problem has recently raised a significant amount of interest because it may provide an important toy model to benchmark genuine quantum relativistic effect like the Zitterbewegung, in the so-called quantum simulation of Dirac equation  \cite{Geritsima}. The equation  can be written in the two component form  \cite{Longhi} as follows
\begin{equation}\label{EQ36}
i \hbar \partial_{t} \underline{\Psi} = c \hat{\alpha} p \underline{\Psi} + \hat{\beta} m c^{2} \underline{\Psi}.
\end{equation}
In the dimensionless coordinates $\tau$ and $\xi$ it can be expressed by
\begin{equation}\label{EQ37}
i \partial_{\tau} \underline{\Psi} = -i \hat{\alpha} \partial_{\xi} \underline{\Psi} + \hat{\beta} \underline{\Psi},
\end{equation}
where
$\hat{\alpha}$ and $\hat{\beta}$ are Pauli-like matrices satisfying the identities
\begin{equation}\label{EQ38}
\hat{\alpha}^{2} = \hat{\beta}^{2}=\hat{1} = \left(\!\begin{array}{cc}
1 & 0\\
0 & 1
\end{array}\!\right), \qquad
\hat{\alpha}\hat{\beta}+\hat{\beta}\hat{\alpha} = 0
\end{equation}
and numerically given as
\begin{equation}\nonumber
\hat{\alpha} = \left(\!\begin{array}{cc}
0 & 1 \\
1 & 0
\end{array}\!\right) \quad\text{and}\quad \hat{\beta} = \left(\!\begin{array}{cc}
1 & 0 \\
0 & -1
\end{array}\!\right).
\end{equation}
We have denoted by $\underline{\Psi} = \left(\!\psi_{+} \atop \psi_{-}\!\right)$ the two component vector, replacing the wave function $\Psi$. The vector $\underline{\Psi}$ should not be confused with a spinor and Eq. \eqref{EQ37} is not the relativistic counterpart of the Pauli equation since we are not dealing with a problem including the spin degrees of freedom. The reason for using two dimensional matrices is that we want Eq. \eqref{EQ37} to describe negative and positive energy states. 

\begin{figure}[!h]
\includegraphics[scale=0.4]{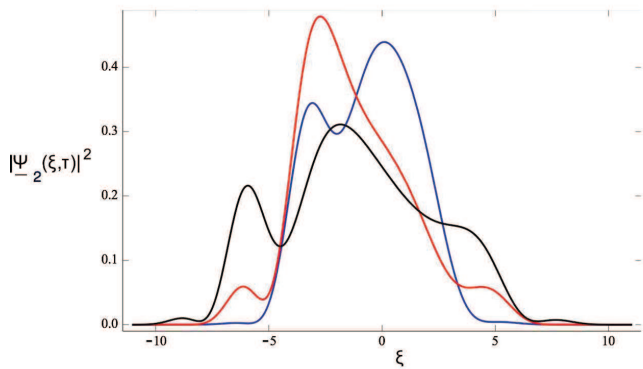}%{fig4b.png}
\caption{\label{rys3}Evolution of an initial wave packet with both energy components as in Eq. \eqref{EQ44} for $\beta = 1$ and different $\tau$ values: $\tau = 4.5$ (blue line), $\tau = 6.5$ (red line), and $\tau = 8.5$ (black line).}
\end{figure}

The solution of Eq. \eqref{EQ37} can be formally obtained, using the evolution operator method, as 
\begin{equation}\label{EQ39}
\underline{\Psi}\left(\xi,\tau\right) = \hat{U}(\tau)\underline{\Psi}(\xi, 0)  
\end{equation}
with the evolution operator written in the form of the $2\times 2$ matrix \cite{Babusci11b}
\begin{align}\label{EQ40}
\hat{U}(\tau) & = \exp\left[\tau\!\left(\!\begin{array}{cc}
i & \partial_{\xi}\\
\partial_{\xi} & -i
\end{array}\!\right)\!\right] \nonumber \\[0.6\baselineskip]
&= \left(\!\begin{array}{cc}
i \hat{A}(\tau) + \hat{B}(\tau) & \hat{A}(\tau) \partial_{\xi} \\
\hat{A}(\tau) \partial_{\xi} & -i \hat{A}(\tau) + \hat{B}(\tau)
\end{array}\!\right),
\end{align}
where $\hat{A}(\tau) = \sin(\sqrt{\hat{\Delta}} \tau)/\sqrt{\hat{\Delta}}$, $\hat{B}(\tau) = \cos(\sqrt{\hat{\Delta}}\tau)$, and $\hat{\Delta}=1-\partial_{\xi}^{2}$. This evolution operator is unitary and therefore the Dirac wave function remains normalized at any time, namely
\begin{equation}\label{EQ41}
\int_{-\infty}^{\infty}\!\! |\underline{\Psi}(\xi, \tau)|^{2} d\xi = \int_{-\infty}^{\infty}\!\! (|\psi_{+}(\xi, \tau)|^{2} + |\psi_{-}(\xi, \tau)|^{2})\, d\xi = 1.
\end{equation}
The use of the Fourier transform method allows one to obtain the solution of the free particle Dirac equation: from Eq. \eqref{EQ39} with $\hat{U}(\tau)$ given in Eq. \eqref{EQ40} we get the evolution of the two components in the form
\begin{align}\label{EQ42}
\psi_{+}(\xi, \tau) & = \int_{-\infty}^{\infty} \left\{i\, \frac{\sin(\sqrt{1+k^{2}}\, \tau)}{\sqrt{1+k^{2}}} [\tilde{\psi}_{+}(k, 0) + k \tilde{\psi}_{-}(k, 0)] \right.\nonumber \\
& \left. + \cos(\sqrt{1+k^{2}}\, \tau)\, \tilde{\psi}_{+}(k, 0)\right\} e^{ik\xi} \frac{dk}{\sqrt{2\pi}},\\
\psi_{-}(\xi, \tau) & = \int_{-\infty}^{\infty} \left\{i\, \frac{\sin(\sqrt{1+k^{2}}\, \tau)}{\sqrt{1+k^{2}}} [k\tilde{\psi}_{+}(k, 0) - \tilde{\psi}_{-}(k, 0)] \right. \nonumber \\
&\left. + \cos(\sqrt{1+k^{2}}\, \tau)\, \tilde{\psi}_{-}(k, 0)\right\} e^{ik\xi} \frac{dk}{\sqrt{2\pi}}, \label{EQ43}
\end{align}
where $\tilde{\psi}_{\pm}(k, 0)$ are the Fourier transforms of $\psi_{\pm}(\xi, 0)$. 
In  Fig. \ref{rys3} we present the snapshots of the time evolution for  $|\Psi_{1}(\xi, \tau)|^2$, calculated using Eqs. \eqref{EQ41},  \eqref{EQ42}  and \eqref{EQ43},  with the initial wave packet being a Gaussian containing both positive and negative energy components  
\begin{equation}\label{EQ44}
\tilde{\underline{\Psi}}_{1}(k, 0) = e^{-\beta k^{2}} \left(1 \atop 1\right)
\end{equation}
 
The presence of the negative energy term yields the behaviour with some features not exhibited by the Salpeter equation. These effects are due to the superposition between the two states, which induces also the characteristic trembling motion of the packet average position and yielding the ``Zitterbewegung'' \cite{Wang}, namely an oscillating motion around a linear trajectory. We remark that using Eq. \eqref{EQ42} with the initial condition possessing only the positive energy term which is given by Gaussian function, we get the evolution which is not dissimilar from the one described with the relativistic Schr\"{o}dinger equation and from the behaviour obtained in Ref. \cite{Babusci11}, using different solution techniques.

Following \cite{Kowalski} we can also find an exact solution for the one dimensional massive Dirac equation with the initial condition
\begin{equation}\label{EQ45}
\underline{\tilde{\Psi}}_{2}(k, 0) = e^{-\beta\sqrt{1+k^{2}}} \binom{\psi_{+}(k, 0)}{\psi_{-}(k,0)}. 
\end{equation}
According to Eqs. \eqref{EQ18}, \eqref{EQ42} and taking the integral representation of $(1+k^{2})^{-1/2}$, i.e. $\frac{1}{\sqrt{1+k^{2}}} = \int_{0}^{\infty} e^{-s\sqrt{1+k^{2}}} ds$, we obtain
\begin{align}\label{EQ46}
\psi_{+}(\xi, \tau) & = \frac{1}{\sqrt{2\pi}}\left[K_{0}(c^{\star}) + \frac{\beta-i\tau}{c^{\star}} K_{1}(c^{\star}) \right] \nonumber \\
& + \frac{1}{\sqrt{2\pi}}\left[- K_{0}(c) + \frac{\beta+i\tau}{c} K_{1}(c) \right]
\end{align}
and
\begin{equation}\label{EQ47}
\psi_{-}(\xi, \tau) = \frac{1}{\sqrt{2\pi}} \left[\frac{\xi}{c} K_{1}(c) - \frac{\xi}{c^{\star}} K_{1}(c^{\star})\right],
\end{equation}
where $c = \sqrt{(\beta+i\tau)^{2} + \xi^{2}}$ and $c^{\star}$ is its complex conjugation.

\section{One Dimensional Klein-Gordon Equation}

The free particle Klein Gordon (FP-KG) equation \cite{Baym} has played a fundamental role in the description of relativistic spin 0 particles. It is well known that according to Feshback Villars transformation \cite{Feshbach}, FP-KG is transformed in two component Schr\"{o}dinger like equation which exhibits a Hamiltonian with unconventional properties. In this Section we treat this transformation from a different perspective. We will look for a formal solution of the free particle KG equation by treating it as a D'Alembert equation and next we will show that the solution of the 1-dimensional free particle KG can be expressed in terms of those of the Dirac equation.

Let FP-KG be written in the dimensionless coordinates $\xi$ and $\tau$ introduced in Sec. III as follows
\begin{equation}\label{EQ48}
\partial_{\tau}^{2} \Psi(\xi, \tau) = - (1 - \partial_{\xi}^{2})\Psi(\xi, \tau)
\end{equation}
with the initial conditions
\begin{equation}\label{EQ49}
\Psi(\xi, 0) = \Psi_{0}(\xi),\qquad \left[\partial_{\tau}\Psi(\xi, \tau)\right]_{\tau=0} = \Psi_{1}(\xi).
\end{equation}
Using the formal procedure presented in the previous Sections for solving Eqs. \eqref{EQ48} and \eqref{EQ49} we can cast our problem in the form of an ordinary second order equation with constant coefficients, namely,
\begin{equation}\label{EQ50}
\partial_{\tau}^{2}\Psi(\xi, \tau) = - \hat{D}^{2}\Psi(\xi, \tau), \quad \hat{D} = (1-\partial_{\xi}^{2})^{1/2},
\end{equation}
which is resembling that of a harmonic oscillator. Therefore, we postulate that the solution of free particle KG can be written in the form
\begin{equation}\label{EQ51}
\Psi_{KG}(\xi, \tau) = e^{i\tau\hat{D}} C_{1}(\xi) + e^{-i\tau\hat{D}} C_{2}(\xi),
\end{equation}
where $C_{1,2}$ are linked to the initial conditions of Eq. \eqref{EQ49} by the identities
\begin{equation} \label{EQ52}
C_{1} + C_{2} = \Psi_{0}(\xi), \qquad C_{1} - C_{2} = -i \hat{D}^{-1} \Psi_{1}(\xi).
\end{equation}
Thus, 
inserting the explicit formulas on $C_{1}$ and $C_{2}$ into Eq. \eqref{EQ51} and making simple symbolic manipulations, we get
\begin{align}\label{EQ53}
\Psi_{KG}(\xi, \tau) & = \ulamek{1}{2} [{_{1}\Psi_{R-S}}(\xi, \tau) + {_{2}\Psi_{R-S}}(\xi, -\tau)] \\
&  = \cos(\tau \hat{D})\Psi_{0}(\xi) + \sin(\tau \hat{D}) \hat{D}^{-1}\Psi_{1}(\xi). \label{EQ54}
\end{align}
The function $_{1}\Psi_{R-S}\left(\xi,\tau\right)$ is the solution of the (forward) free relativistic Schr\"{o}dinger particle with initial condition $\Psi_{0}(\xi) - i\hat{D}^{-1}\Psi_{1}(\xi)$ and $_{2}\Psi_{R-S}(\xi, -\tau)$ is its backward counterpart, with initial condition $\Psi_{0}(\xi) + i\hat{D}^{-1}\Psi_{1}(\xi)$. 

\begin{figure}
\begin{center}
\includegraphics[scale=0.42]{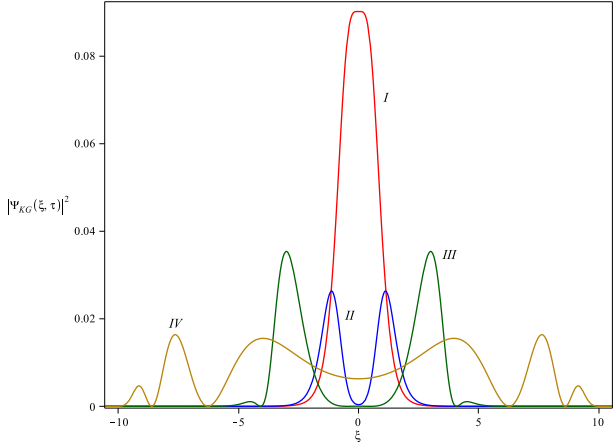}
\caption{\label{7Jul14-1} Evolution of the FP-KG solution, Eq. \eqref{EQ55}, at different times: $\tau=0.5$ (I; red line), $\tau = 1$ (II; blue line), $\tau = 4$ (III; green line),  and $\tau = 9.5$ (IV; gold line).}
\end{center}
\end{figure}

The evolution of the solution of the free particle KG equation is given in Fig. \ref{7Jul14-1}, where we have used an initially Macdonald distribution, namely for $\tilde{\Psi}_{0}(k) = e^{-\beta\sqrt{1+k^{2}}}$ and $\tilde{\Psi}_{1}(\xi) = 0$. Then, Eq. \eqref{EQ54} gives
\begin{equation}\label{EQ55}
\Psi_{KG}(\xi, \tau) = \left\{\frac{\beta-i\tau}{\sqrt{2\pi} c^{\star}} K_{1}(c^{\star}) + \frac{\beta+i\tau}{\sqrt{2\pi} c} K_{1}(c) \right\},
\end{equation}
where $c$ and its complex conjugation $c^{\star}$ are defined below Eq. \eqref{EQ47}. This is the effect of the interference with the backward time solution which yields a fairly rich evolution according to which the wave function undergoes a free propagation. The use of a Gaussian does not provide any significant difference.

The evolution operator associated with Eq. \eqref{EQ48}, related to two components Schr\"{o}dinger-like equation, should be written as the matrix $2\times 2$:
\begin{equation}\label{EQ56}
\hat{U}_{KG}(\tau) = \left(\begin{array}{cc} \cos(\tau\hat{D}) & \sin(\tau\hat{D})\hat{D}^{-1} \\
-\sin(\tau\hat{D})\hat{D} & \cos(\tau\hat{D})\end{array}
\right),
\end{equation}
whose determinant is equal to one. The inverse operator of $\hat{U}_{KG}(\tau)$ exists and it fulfils the relation $[\hat{U}_{KG}(\tau)]^{-1} = \hat{U}_{KG}(-\tau)$. In consequence Eq. \eqref{EQ48} can be rewritten as
\begin{equation}\label{EQ57}
\left({\Psi_{0}(\xi, \tau) \atop \Psi_{1}(\xi, \tau)}\right) = \hat{U}_{KG}\left({\Psi_{0}(\xi) \atop \Psi_{1}(\xi)}\right),
\end{equation}
where the equation on the upper component $\Psi_{0}(\xi, \tau)$ is given in Eq. \eqref{EQ54}, whereas $\Psi_{1}(\xi, \tau) = \partial_{\tau} \Psi_{KG}(\xi, \tau)$. Moreover, the conserved quantity associated with such an evolution operator is $\int_{-\infty}^{\infty}(|\Psi_{0}(\xi)|^{2} + \hat{D}^{-2}|\Psi_{1}(\xi)|^{2}) d\xi$; which cannot be understood in terms of probability rather it has been constructed in analogy with the energy of an harmonic oscillator. 

Let us observe that in Fourier space Eq. \eqref{EQ54} reads 
\begin{equation}\label{EQ58}
\tilde{\Psi}_{KG}(k, \tau) = \cos[\tau\omega(k)] \tilde{\Psi}_{0}(k) + \frac{\sin[\tau\omega(k)]}{\omega(k)} \tilde{\Psi}_{1}(k),
\end{equation}
where $\omega(k) = \sqrt{1+k^{2}}$ and $\tilde{\Psi}_{i}(k)$, $i=0,1$ are the initial condition written in the Fourier space. The analogy with harmonic oscillator is made more clear by noting that in the Fourier space, the quantity
\begin{equation}\label{EQ59}
\int_{-\infty}^{\infty} [\partial_{\tau} \tilde{\Psi}_{KG}(k, \tau)]^{2} + [\omega(k) \tilde{\Psi}_{KG}(k, \tau)]^{2} dk 
\end{equation}
is a constant of motion.

Finally, we point out that it is also possible express the solution of our problem in terms of the free particle Dirac solution by setting 
\begin{equation}\label{EQ60}
\hat{D} = -i \hat{\alpha} \partial_{\xi} + \hat{\beta}.
\end{equation}

\section{Quantum Relativistic Particles in a Linear Potential}

\subsection{The Salpeter equation with a linear potential}

Let us consider the Salpeter equation with a potential function $V(\xi)$
\begin{align}\label{EQ61}
i\partial_{\tau}\Psi(\xi, \tau) & = \sqrt{1-\partial_{\xi}^{2}} \Psi(\xi, \tau) + \mu_{0} V(\xi),\nonumber \\
\Psi(\xi, 0) & = \Psi_{0}(\xi),
\end{align}
which in the Fourier space reads
\begin{equation}\label{EQ62}
i\partial_{\tau}\tilde{\Psi}(k, \tau) = \sqrt{1+k^{2}} \tilde{\Psi}(k, \tau) + \mu_{0}\tilde{V}\tilde{\Psi}(k, \tau),
\end{equation}
where $\tilde{V} = V(i\partial_{k})$ and $\mu_{0}$ is a dimensionless constant. We note that, in general case, the presence of an operator $\tilde{V}$, which doesn't commute with the kinetic term, hampers the possibility of a straightforward analytical solution for Eq. \eqref{EQ62}, which can however be obtained for $V(\xi)$ being a linear function whose physically meaningful example is when a relativistic, charged particle interacts with a static constant electric field. In this case, Eq. \eqref{EQ62} can be rewritten in the form
\begin{equation}\label{EQ63}
i\partial_{\tau}\tilde{\Psi}(k, \tau) = \sqrt{1+k^{2}} \tilde{\Psi}(k, \tau) + i\mu_{0}\partial_{k}\tilde{\Psi}(k, \tau).
\end{equation}
To solve Eq. \eqref{EQ63} first we eliminate the derivative by the use of the transformation
\begin{equation}\label{EQ64}
\tilde{\Psi}(k, \tau) = e^{\tau\mu_{0}\partial_{k}}\tilde{\Phi}(k, \tau),
\end{equation}
which yields a modified equation:
\begin{align}\label{EQ65}
i\partial_{\tau}\tilde{\Phi}(k, \tau) &= \hat{K}\tilde{\Phi}(k, \tau) \\
\hat{K}(k, \tau) &= e^{-\tau\mu_{0}\partial_{k}} \sqrt{1+k^{2}} e^{\tau\mu_{0}\partial_{k}}. \label{EQ66} 
\end{align}
The use of the procedure touched upon in Ref. \cite{Babusci13} allows to cast its formal solution in the form
\begin{equation}\label{EQ67}
\tilde{\Phi}(k, \tau) = \exp\left[ -i\int_{0}^{\tau}\!\! \hat{K}(k, \tau') d\tau' \right] \tilde{\Phi}_{0}\left(k\right)
\end{equation}
where $\tilde{\Phi}_{0}(k)\equiv \tilde{\Psi}_{0}(k)$. The solution can finally be obtained in the form
\begin{align}\label{EQ68}
\tilde{\Psi}(k, \tau) &= e^{\tau\mu_{0}\partial_{k}} \tilde{\Phi}(k, \tau) = \tilde{\Phi}(k + \mu_{0}\tau, \tau) \nonumber \\
& = e^{-i\int_{0}^{\tau}\!\! \sqrt{1 + (k + \mu_{0}\tau')^{2}} d\tau'} \tilde{\Psi}_{0}(k + \mu_{0}\tau).
\end{align}

\begin{figure}[!h]
\includegraphics[scale=0.53]{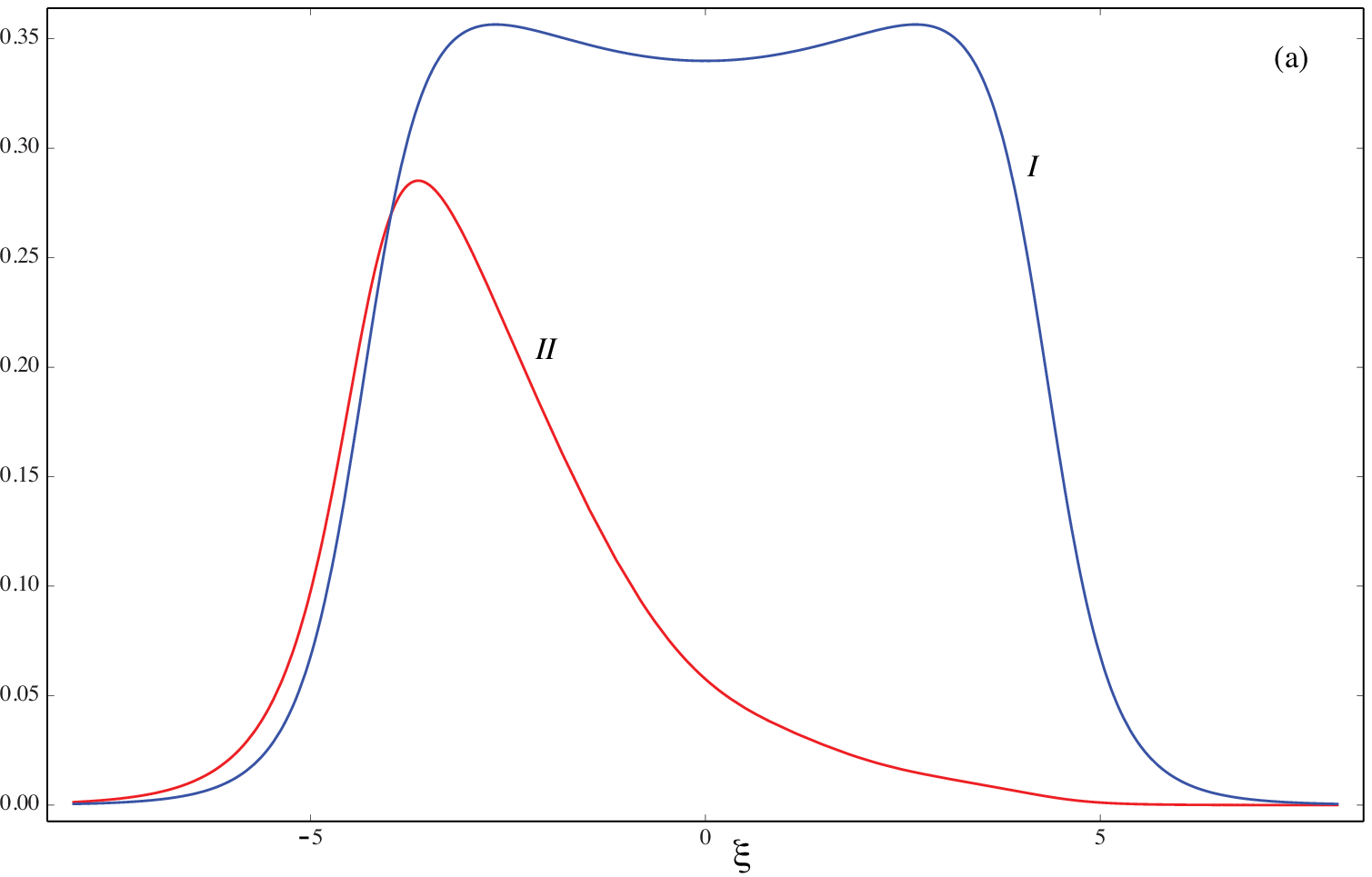}
\includegraphics[scale=0.8]{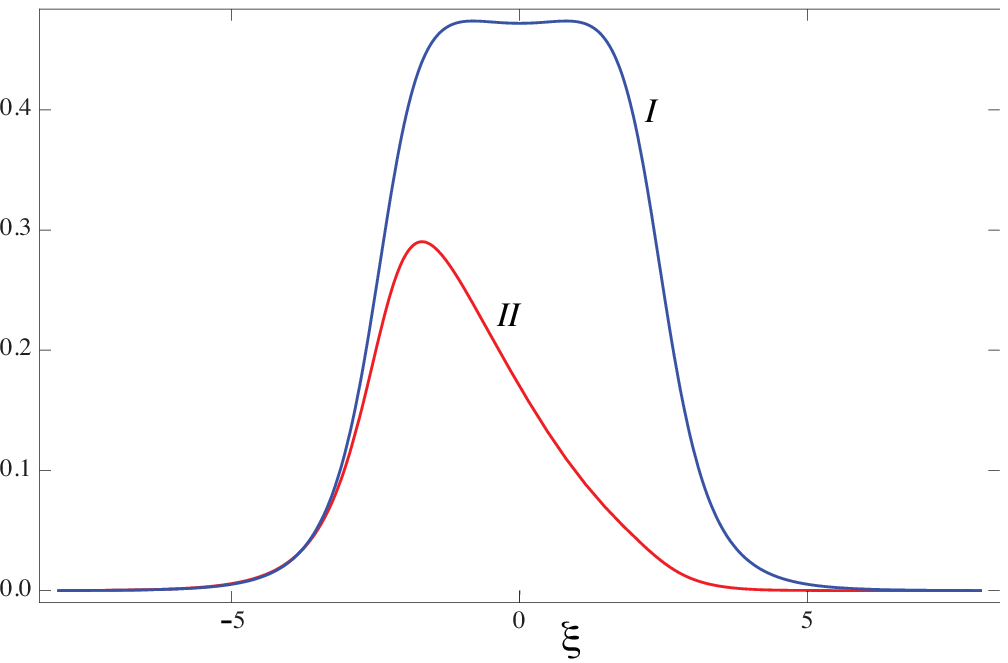}
\caption{\label{fig10} (a) Comparison between modulus of wave function evolution in presence of linear potential (I; blue curve; Eq. \eqref{EQ68} with $\tilde{\Psi}(k)$ given in Eq. \eqref{EQ16}) and the wave function evolution of free particle (II; red curve; Eq. \eqref{EQ27}) at $\tau=4.3$ , (b) the same as (a) but $\tau=2.3$ .}
\end{figure}

In Figs. \ref{fig10} we report a comparison between the solutions for the free particle case and that of  Eq. \eqref{EQ61} with a linear potential put in. In both cases we have assumed as initial wave function the function given in  Eq. \eqref{EQ16} with $\beta = 1$. It is important to emphasize that the presence of the potential  provides a suppression of the emergence of the  bi-modal behaviour by inducing a kind of external localization. 

\subsection{The Dirac equation with a linear potential}

The one dimensional Dirac equation with a linear potential has the form
\begin{equation}\label{EQ69}
i\partial_{\tau}\underline{\Psi} = -i\hat{\alpha} \partial_{\xi} \underline{\Psi} + \hat{\beta}\underline{\Psi} + \xi\hat{\mu}_{0} \underline{\Psi},
\end{equation}
where $\hat{\mu}_{0} = \mu_{0}\left(\begin{array}{c c} 1 & 0 \\ 0 & 1\end{array}\right)$. The solution of the problem is slightly more complicated than in the case of the Salpeter equation since the Hamiltonian is now expressed as linear combination of Pauli matrices. The formal solution of Eq. \eqref{EQ69} can be given by using the operational method with the evolution operator defined as
\begin{equation}\label{EQ70}
\hat{U} = e^{\tau\hat{\alpha} \partial_{\xi} + i\xi \tau \hat{\mu}_{0} + i\tau \hat{\beta}}.
\end{equation}
Albeit the problem is amenable for an exact treatment, we use the Zassenhaus formula at the second order in the commutator expansion  \cite{Bab_Dat_Del} and write the operator $\hat{U}$  in the following disentangled form
\begin{equation}\label{EQ71}
\hat{U} = e^{i\xi\hat{\mu}_{0}\tau + i \tau \hat{\beta}} e^{\tau\hat{\alpha}\partial_{\xi}} e^{i \ulamek{\tau^{2}}{2} \mu_{0} \hat{\alpha}},
\end{equation}
which is valid at the order $O(t^{3})$. The repeated application of the evolution operator \eqref{EQ71} on the initial state, yields the solution in the form of the following recursion
\begin{align}\label{EQ72}
\underline{\Psi}_{n}(\xi, \tau) & = \left(\begin{array}{cc}
e^{i \tau (1 + \mu_{0}\xi)} & 0\\
0 & e^{-i \tau (1 - \mu_{0}\xi)}
\end{array}\right) \nonumber\\
&\times
\left(\begin{array}{cc}
\cosh(\tau \partial_{\xi}) & \sinh(\tau \partial_{\xi})\\
\sinh(\tau \partial_{\xi}) & \cosh(\tau \partial_{\xi})
\end{array}\right)
\underline{\Psi}_{n-1}(\xi, \tau), 
\end{align}
where
\begin{align}\label{EQ73}
\underline{\Psi}_{0}(\xi, \tau) & =\left({\psi_{+}(\xi) \atop \psi_{-}(\xi)}\right)_{0}, \\
\underline{\Psi}_{1}(\xi, \tau) & = \left(
{[\psi_{c}^{+}(\xi) + \psi_{s}^{-}(\xi)] e^{i\tau (1 - \mu_{0}\xi)}\atop 
[\psi_{c}^{-}(\xi) + \psi_{s}^{+}(\xi)] e^{-i\tau (1 + \mu_{0}\xi)} }
\right)_{0}, \label{EQ74}
\end{align}
with 
\begin{equation}\label{EQ74}
\psi_{c, s}^{\pm}(\xi) = \frac{1}{2} [\psi_{\pm}(\xi + \tau) \pm \psi_{\pm}(\xi - \tau)].
\end{equation}

The previous iteration can be extended to any type of potential and specific relevant examples will be discussed in a paper specially dedicated to this problem.

\section{Conclusion}

In this paper we have presented a number of issues also of practical nature to deal with problems concerning relativistic equations and other fractional evolution problems often encountered in physics. We have left open many points and in particular the link of the present treatment with previous and well established methods. In particular the use of Green function technique for the treatment of the Klein Gordon equation or the method associated with the use of Wightman solution    \cite{Derezinski} . In this last case the solution of the free particle Klein Gordon writes, in terms of Wightman (and anti-Wightman as well) functions expressed as combinations of Hankel and Macdonald functions of the first kind.
The relevant link with our formalism and with other type of solutions (Pauli-Jordan, Feynman, Stueckelberg) will be the topic of a forthcoming investigation. We have touched on many aspects of a relevant formalism and we have fixed the main mathematical apparatus which can be exploited to discuss a number of problems in relativistic nonlocal quantum mechanics. 

\section{Acknowledgments}

KAP and KG acknowledge support from the PHC Polonium, Campus France, project no. 28837QA. KG thanks support from MNiSW, "Iuventus Plus 2015-2016", program no IP2014 013073.

%-----------------------------------------------------------


\begin{thebibliography}{99}

\bibitem{Baym} G. Baym, \textit{Lectures On Quantum Mechanics} (Benjamin, New York, 1969).

\bibitem{Wilcox} R. M. Wilcox, J. Math. Phys. \textbf{8}, 962 (1967).

\bibitem{Dattoli88} G. Dattoli, J. C. Gallardo and A. Torre, Riv. Nuovo Cimento, \textbf{11} (1988).

\bibitem{Dattoli97} G. Dattoli, P. L. Ottaviani, A. Torre, and L. Vazquez, Riv. Nuovo Cimento \textbf{20}, 1 (1997).

\bibitem{Blanes} S. Blanes, F. Casas, J. A. Oteo and J. Ros, Phys. Rep. \textbf{470}, 151238 (2009).

\bibitem{Glaeske} H.-J. Glaeske, A. P. Prudnikov, and K. A. Sk\'ornik, \textit{Operational Calculus and Related Topics} (Chapman and Hall/CRC, Boca Raton, 2006).
  
\bibitem{Eliazar} I. Eliazar and M. F. Shlesinger, Phys. Rep. \textbf{527}, 101 (2013).

\bibitem{Mark} M. M. Meerschaert, \textit{Fractional Calculus, Anomalous Diffusion, and Probability}, World Scientific Publishing Co. Pte. Ltd. \\http://www.worldscibooks.com/physics/8087.html, chapter 11 (2012).

\bibitem{Klaffter} J. Klafter, M. F. Schlesinger and G. Zumofen, Phys. Today \textbf{49}, 33 (1996).

\bibitem{Barndorff} O. E. Barndorff-Nielsen and N. Shepard, Scand. J. Stat. \textbf{24}, 1 (1997).

\bibitem{Stollenwerk} N. Stollenwerk and J. P. Boto, \textit{Fractional calculus and L\'{e}vy fl{}ights: modelling spatial epidemic
spreading}, Proceedings of the International Conference on Computational and Mathematical Methods in Science and Engineering,
CMMSE 2009 30 June, 1(2009).

\bibitem{Gonzalez} M. C. Gonzalez, C. A. Hidalgo and A. L. Barabasi, Nature \textbf{453}, 779 (2008).

\bibitem{Ramos} G. Ramos-Fernandez et al., Behav. Ecol. Sociobiol.  \textbf{273}, 1743 (2004).

\bibitem{Sims} D. W. Sims et al., Nature \textbf{451}, 1098 (2008).

\bibitem{Babusci11} D. Babusci, G. Dattoli, and M. Quattromini, Phys. Rev. A \textbf{83}, 062109 (2011).

\bibitem{Babusci13} D. Babusci, G. Dattoli, M. Quattromini, and E. Sabia, Phys. Rev. E \textbf{87}, 033202 (2013).

\bibitem{GreinerBjoerken} W. Greiner, \textit{Relativistic Quantum Mechanics - Wave Equations}, 3rd Ed., (Springer, Berlin, 2000); J. D. Bjoerken, and S. D. Drell, \textit{Relativistic Quantum Mechanics} (McGraw-Hill, New York, 1964).

\bibitem{KowalskiRembielinski} K. Kowalski, and J. Rembielinski, Phys. Rev. A \textbf{81}, 012118 (2010).

\bibitem{Strange} P. Strange, \textit{Relativistic Quantum Mechanics with Applications in Condensed Matter and Atomic Physics}  (Cambridge University Press, Cambridge, 1998)

\bibitem{Doetsch} G. Doetsch, \textit{Handbuch der Laplace Transformation} (Birkh\"{a}user, Basel, 1950-1956).

\bibitem{Montroll} E. W. Montroll and J. T. Bendler, J. Stat. Phys. \textbf{34}, 129 (1984).

\bibitem{Penson} K. A. Penson and K. G\'{o}rska, Phys. Rev. Lett. \textbf{105}, 210604 (2010).

\bibitem{Gorska} K. G\'{o}rska and K. A. Penson, Phys. Rev. E \textbf{16}, 061125 (2011), H. Pollard, Bull. Amer. Math. Soc. \textbf{52}, 908 (1946).

\bibitem{Dattoli} G. Dattoli, S. Khan, and P. E. Ricci, Integr. Transf. Spec. Fun. \textbf{19}, 1 (2008) and references therein.

\bibitem{Andrews} L. C. Andrews, \textit{Special Functions for Applied Mathematicians and Engineers} (Mac Millan, New York, 1985).

\bibitem{Prudnikov} A. P. Prudnikov, Yu. A. Brychkov, and O. I. Marichev, \textit{Integrals and Series, vol. 1: Elementary Functions} (Gordon and Breach,
Amsterdam, 1998). 

\bibitem{Kowalski} K. Kowalski and J. Rembielinski, Phys. Rev. A \textbf{84}, 012108 (2011).

\bibitem{Geritsima}  R. Geritsima, G. Kirchmair, F. Zahringer, E. Solano, R. Blatt and C. F. Ross, Nature \textbf{463}, 68 (2010).

\bibitem{Longhi} S. Longhi, Opt. Lett. \textbf{35}, 235 (2010).

\bibitem{Babusci11b} D. Babusci, G. Dattoli, M. Quattromini, and P. E. Ricci, App. Math. Comput. \textbf{218}, 1495 (2011).

\bibitem{Wang} Z.-Y. Wang and C.-D. Xiong, Phys. Rev. A \textbf{77}, 045402 (2008).

\bibitem{Feshbach} H. Feshbach and F. Villars, Rev. Mod. Phys. \textbf{30}, 24 (1958).

\bibitem{Bab_Dat_Del} D. Babusci, G. Dattoli, and M. Delfranco, Lectures
on Mathematical Methods for Physics, Internal Report ENEA RT/2010/5837.

\bibitem{Derezinski} see e. g. Jan Derezi\'{n}ski, Mathematical Introduction to Quantum Field Theory, www.fuw.edu.pl/derezins/


\end{thebibliography}
\end{document}